\documentstyle[12pt,psfig]{article}
\newcommand{\boldarrayrulewidth}{1pt} 

\def\bhline{\noalign{\ifnum0=`}\fi\hrule \@height  
\boldarrayrulewidth \futurelet \@tempa\@xhline}

\def\@xhline{\ifx\@tempa\hline\vskip \doublerulesep\fi
      \ifnum0=`{\fi}}

\newcommand{\m}{\medbreak}
\newcommand{\no}{\noindent}
\newcommand{\EQ}{\begin{equation}}
\newcommand{\eq}{\end{equation}}
\newcommand{\EQA}{\begin{eqnarray}}
\newcommand{\eqa}{\end{eqnarray}}

\newcommand{\ALLPV}{\mbox{$A_{LL}^{PV}\ $}}

\def\pr#1#2#3{ Phys. Rev. {\bf{#1}} (#2) #3}
\def\prl#1#2#3{ Phys. Rev. Lett. {\bf{#1}} (#2) #3}
\def\pl#1#2#3{ Phys. Lett. {\bf{#1}} (#2) #3 }

\def\np#1#2#3{ Nucl. Phys. {\bf{#1}} (#2) #3}
\def\zp#1#2#3{ Z. Phys. {\bf{#1}} (#2) #3}

\begin{document}
\begin{titlepage}
\vspace{0.2in}
\vspace*{1.5cm}
\begin{center}
{\large \bf Search for compositeness and leptophobic gauge bosons \\
(with polarized beams)
\\} 
\vspace*{0.8cm}
{\bf J.-M. Virey}{$^1$}  \\ \vspace*{1cm}
Centre de Physique Th\'eorique$^{\ast}$, C.N.R.S. - Luminy,
Case 907\\
F-13288 Marseille Cedex 9, France\\ \vspace*{0.2cm}
and \\ \vspace*{0.2cm}
Universit\'e de Provence, Marseille, France\\
\vspace*{1.8cm}
{\bf Abstract \\}
\end{center}
In a first step, we explore the discovery and analysis potentials of the
HERA collider, with and without
polarized beams, in the search for electron-quark 
compositeness in the neutral current channel. 
Then we study the parity violating effects, for jet
production in polarized $pp$ collisions at RHIC, 
which could be due to the presence
of quark subconstituents or new massive gauge bosons. We emphasize that the measurement of 
spin asymmetries in such a polarized context could give some crucial informations
on the chiral structure of these hypothetical new interactions.
\vspace{0.5cm}

\begin{center}
{\it To appear in the proceedings of the Workshop "Beyond the Desert", Ringberg,
Germany, June 1997.
}
\end{center}

\vfill
\begin{flushleft}
\m\no
Number of figures : 2\\

\m\no
July 1997\\
CPT-97/P.3514\\
\m\no
anonymous ftp or gopher : cpt.univ-mrs.fr

------------------------------------\\
$^{\ast}$Unit\'e Propre de Recherche 7061

{$^1$} Moniteur CIES and allocataire MESR \\
E-mail : Virey@cpt.univ-mrs.fr
\end{flushleft}
\end{titlepage}

\section{Electron-quark compositeness at HERA}
The idea of compositeness has been introduced in the hope of solving some of the
problems of the Standard Model (SM). For instance, it could explain the 
generation pattern of the SM, or, like the technicolor idea,
represent an alternative to the 
scalar sector of the theory. The phenomenological approch is to consider 
a new contact interaction
between electron and quark subconstituents,
which is normalized to a certain
compositeness scale $\Lambda$. This is represented by the following effective 
lagrangian \cite{ELP83,Ruckl84}:
\EQA\label{Leq}
{\cal L}_{eq} = \sum_q ( \eta^q_{LL}(\bar e_L \gamma_{\mu} e_L) 
(\bar q_L \gamma^{\mu} q_L) +  \eta^q_{RR}(\bar e_R \gamma_{\mu} e_R) 
(\bar q_R \gamma^{\mu} q_R)  \nonumber 
  \\   + \eta^q_{LR}(\bar e_L
\gamma_{\mu} e_L)  (\bar q_R \gamma^{\mu} q_R) + \eta^q_{RL}(\bar e_R
\gamma_{\mu} e_R)  (\bar q_L \gamma^{\mu} q_L) )  
\eqa

\no with $\eta^q_{ij}= \epsilon \frac{g^2}{(\Lambda^q_{ij})^2}$ where
$g^2=4 \pi$ and $\epsilon=\pm 1$. The sign $\epsilon$ caracterises the 
nature of the interferences with SM amplitudes. 
The four subscripts $LL$, $RR$, $LR$ and $RL$ caracterise the chiral
structure of the new interaction. 
These four chiralities, along with the sign $\epsilon$, define 8 individual
models. The composite interaction could correspond to one of these models or
to any combination of them.

We know from atomic parity violation (APV) experiments on Cesium atoms that these
individual models are severely constrained, giving some bounds of the order
of $\Lambda \sim 10$ TeV \cite{Deandrea}. However it appears that it is easy to find
some combinations of the chiralities which evade these constraints 
\cite{Nels}. In the following, for simplicity, we will consider the 8 models
individually and then observe the effects of more complicated models at the end
of this section.

Recently, the H1 and ZEUS collaborations at HERA have observed an excess of
events, in comparison with the SM expectations, at high $Q^2$, in the deep
inelastic positron-proton cross section $\sigma_+ \equiv \frac{d\sigma}{dQ^2}
(e^+p \rightarrow e^+ X)$ \cite{HERAe}. It could be interpreted as a manifestation
of electron-quark compositeness \cite{CTe} for a scale $\Lambda \sim 3\, TeV$ in the
up quark sector. Note that, since the lepton beam is made of positrons, the
cross section $\sigma_+$ is sensitive to the chiralities $LR^\pm$ and/or $RL^\pm$
where $\pm$ correspond to $\epsilon$.
Several other possible
explanations have also been mentionned in this conference.

With the present values for the parameters
of the HERA experiment \cite{HERAe} but with higher integrated luminosities
($L_{e^-}=L_{e^+}=1\, fb^{-1}$), we can show \cite{jmvp,pcjmv} that cross section measurements
can probe a compositeness scale of the order of $7\, TeV$ for constructive
interferences ($\epsilon=+1$), and of the order of $6\, TeV$ in the destructive case. 
We deduce that the present HERA anomaly will be soon confirmed or
invalidated. However it appears that electron {\it and} positron beams
are necessary to cover all the possible chiralities of the new interaction.
The comparison of the two cross section $\sigma_-$ and $\sigma_+$
allow the distinction of 2 classes of chiralities : $(LL^\pm,RR^\pm)$
and $(LR^\pm,RL^\pm)$ \cite{martyn,CTe,jmvp}.

Now, we want to emphasize that the measurement of some spin asymmetries,
defined in the context of HERA with polarized lepton beams and also with a
polarized proton beam, could give some very important information
on the chiral structure of the new interaction. Note that lepton
polarization is a part of the HERA program, and that proton polarization is seriously
under discussion \cite{felt}.\\
\no The evaluation of the cross section, asymmetries and their errors
is made with the folowing parameters :
$\sqrt{s} = 300\, GeV$, $L_{e^\pm} = 250\, pb^{-1} $ per spin configuration,
$0.01 < y < 0.95$. Concerning the $Q^2$ resolution we take 
$\frac{\Delta Q^2}{Q^2}
= 34.3\,\%$ and $Q^2_{min} = 200.\, GeV^2$ which defines
16 points of analysis on the energy domain $200\, GeV^2 < Q^2 < 5. 10^4 GeV^2$.
GRSV polarized parton distributions \cite{GRSV} are used for the calculations. 
This choice corresponds to a conservative attitude since for this set 
of distributions the quarks are
weakly polarized in comparison with other sets which are currently used, like GS96 or BS
\cite{param}. As a consequence, the spin effects are weaker giving smaller bounds on 
$\Lambda$.
The degrees of polarization of the beams are taken to
$P_{e^-}=P_{e^+}=P_{p}=70 \, \%$. 
Finaly, we have chosen a total systematical error of 10\% for the asymmetries :
$\frac{\Delta A_{syst}}{A}\; =\; 10\, \%$.
\vspace{0.5cm}

\no \underline{\it Results} \cite{jmvp} :
\vspace{0.4cm}

We have simulated 60 spin asymmetries 
that we can construct with the 8 independent cross sections :
\EQ
\sigma_-^{--} \;\;\;\;\sigma_-^{++} \;\;\;\;\sigma_-^{-+} \;\;\;\;\sigma_-^{+-} 
\;\;\;\;\; and \;\;\;\;\;
\sigma_+^{--} \;\;\;\;\sigma_+^{++} \;\;\;\;\sigma_+^{-+} \;\;\;\;\sigma_+^{+-}
\eq
\no where $\sigma_i^{\lambda_e \lambda_p} \equiv (\frac{d\sigma_t}{dQ^2})^{\lambda_e 
\lambda_p}$, $i$ refers to the electric charge of the colliding lepton and
$\lambda_e, \lambda_p$ are the helicities of the lepton and the proton.

\no It appears that the observables which are the most sensitive to the presence 
of contact interactions, are the parity violating spin asymmetries :
\EQ\label{defALLPV}
A_{LL}^{PV}(e^-)\; =\; \frac{\sigma^{--}_- \, -\, \sigma^{++}_-}{\sigma^{--}_- \, 
+\, \sigma^{++}_-}
\;\;\;\;\;\; and \;\;\;\;\;\;
A_{LL}^{PV}(e^+)\; =\; \frac{\sigma^{--}_+ \, -\, \sigma^{++}_+}{\sigma^{--}_+ \, 
+\, \sigma^{++}_+}
\eq
\no and the "mixed" charge-spin asymmetry :
\EQ\label{defB22}
B_2^2\; =\; \frac{\sigma_-^{++} \, -\, \sigma_+^{++}}{\sigma_-^{++} \, +\, 
\sigma_+^{++}}
\eq
\no Using a $\chi^2$ analysis we obtain the bounds presented in Table 1.

\vspace{0.5cm}

\begin{center}
\begin{tabular}{|c||c|c|c|c||c|c|c|c|}
\hline
$\Lambda$ (TeV) & $\Lambda_{LL}^+$&$\Lambda_{RR}^+$&$\Lambda_{LR}^+$
&$\Lambda_{RL}^+$&$\Lambda_{LL}^-$&$\Lambda_{RR}^-$&$\Lambda_{LR}^-$&$\Lambda_{RL}^-$\\
\hline
${\scriptstyle A_{LL}^{PV}(e^\pm)\, \& \, B_2^2}$& $6.6$ & $7.2$ 
& $7.0$ & $7.0$ & $6.3$ & $7.0$ & $6.8$ & $6.7$ \\
\hline
\end{tabular} 
\end{center}
\begin{center}
Table 1: Limits on $\Lambda$ at 95\% CL.
\end{center}

\no We find that the limits are comparable to the unpolarized case \cite{jmvp,martyn}.
They are slightly better for destructive interferences 
($\epsilon=-1$).

\begin{figure}[ht]
\vspace{-0.5cm}
    \centerline{\psfig{figure=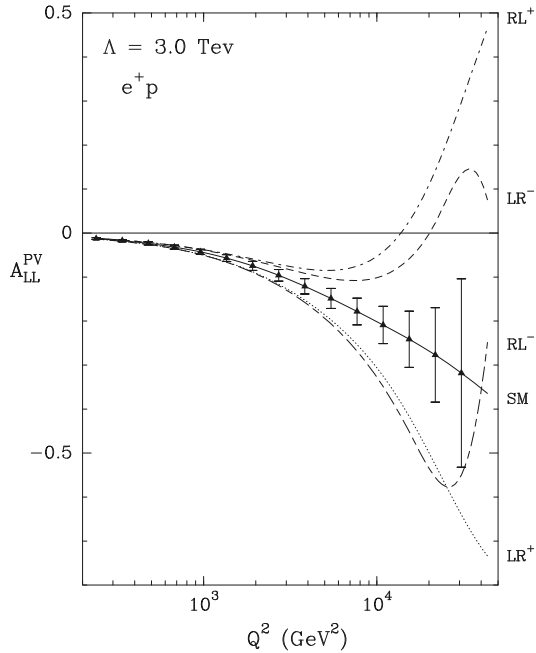,width=8cm}}
\vspace{-1cm}
    \caption[]{The asymmetry $A_{LL}^{PV}(e^+)$. Predictions for
SM (plain curve) and for contact interactions (labels on fig.) with $\Lambda = 3\, TeV$. 
The chiralities which are not mentionned are close to SM.}
    \label{figb1} 
\end{figure}

\no $A_{LL}^{PV}(e^+)$ is represented on Fig.1. This asymmetry is sensitive to the
chiralities ($LR^\pm,RL^\pm$). On the other hand, the direction of the deviation from SM
expectations allows now the distinction between 2 classes of chiralities :
($LR^-,RL^+$) for a positive deviation {\it or} ($LR^+,RL^-$) for a negative one.
Similarly, we can show easily that $A_{LL}^{PV}(e^-)$ is sensitive to the
chiralities ($LL^\pm,RR^\pm$). Then, the same procedure allows to
discriminate between ($LL^+,RR^-$) and ($LL^-,RR^+$). We deduce that the measurement of
these two asymmetries would allow to separate the 4 classes :
\no ($LL^+,RR^-$), ($LL^-,RR^+$), ($LR^+,RL^-$) and ($LR^-,RL^+$).

We can go further in the identification of the chiral structure of the new interaction
by the use of additionnal asymmetries. For instance, $B_2^2$ is strongly sensitive
to the presence of the chiralities ($RR^\pm,LR^\pm$). Again the direction of the 
deviation from SM distinguish ($RR^+,LR^-$) from ($RR^-,LR^+$). But, since these
2 classes are distinct from the 4 precedent ones, we conclude that the measurements
of the three spin asymmetries $A_{LL}^{PV}(e^-)$, $A_{LL}^{PV}(e^+)$ and $B_2^2$
should give a clear identification of the chiral structure of the new interaction
{\it in this na\"{\i}ve model}.

Now, it turns out that
if the chiral structure of the new interaction is more complicated,
in general, the three asymmetries mentionned above will be sufficient to 
identify the precise chiral structure. But for some special cases, 
like for instance the $VV$ model \cite{Ruckl84} which conserves Parity, 
some cancellations occur, then
we need some other spin asymmetries. It appears that the (four) Parity Conserving
spin asymmetries, defined when only the proton spin is flipped (which minimizes
systematical errors), are particularly interesting since they are mainly 
sensitive to only one chirality. The problem of these PC asymmetries is that 
they are less sensitive to new physics than the PV ones. Then, if the
new interaction have a complicated structure, we can obtain valuable informations
at lower value of $\Lambda$ ($\sim 5$ TeV).

To conclude this part, we can make some remarks on the one spin asymmetries defined
when only the lepton beams are polarized. The behaviour of these asymmetries
had been presented some years ago in \cite{Ruckl84}. It appears \cite{jmvp}
that these asymmetries are less sensitive to the presence of new physics
than the double spin asymmetries. Moreover, we can't define as many asymmetries
as in the 2 spin case. For instance, we can't define the Parity Conserving spin asymmetries.
Then, if the structure of the new interaction is complicated, we can loose
the opportunity to identify its chiral structure.

\section{New Physics at RHIC}

Around the year 2000, the RHIC Spin Collaboration (RSC)
will use the RHIC machine as a polarized $pp$ collider.
The center of mass energy will be as high as 500 GeV, the degree
of beam polarization near 70\% and the
luminosity as high as ${\cal L} = 2. 10^{32} \, cm^{-2}.s^{-1}$ which gives after
a few months of run $L=800\, pb^{-1}$ \cite{RSC}. Moreover,
the possibility of accelerating polarized $^3He$ nuclei, which has been discussed
recently \cite{RSCmeeting},
will open some new perspectives since polarized $pn$ collisions will be allowed.
With these figures,
spin asymmetries as small as 1\% should be measurable. For example, the
PC double-spin longitudinal asymmetry $A_{LL}$ in
inclusive one-jet production or in direct photon production will be obtained 
with very small errors,
hence allowing to test the spin structure of QCD.
Adding the measurement of some PV asymmetries
in $W$ and $Z$ boson production, it will be possible to
determine very accurately the various polarized partonic
distributions inside the proton (see \cite{jsjmv} and references therein).
On the other hand, as was noticed some time ago \cite{sapin},
non conventional PV effects in hard-hadron reactions have never
been searched for. Then, we will focus (again) on the 
double  helicity PV asymmetry $A_{LL}^{PV}$ for the inclusive production 
of one jet. The definition of \ALLPV is similar to eq.\ref{defALLPV} but now
$\sigma^{\lambda,\lambda'}$ means the cross
section in a given helicity configuration for the production
of a single jet with transverse energy $E_T$,
integrated over some rapidity interval around $y=0$.

According to the Standard Model, this process
is essentially governed by QCD plus a small contribution
due to electroweak boson exchanges. The latter induces a small
$A_{LL}^{PV} $ which should  be visible at RHIC. This SM effect,
using GRSV distributions,
is represented by the plain curve on the figure 2. The rise with
$E_T$ is due to the increasing importance
of quark-quark scattering relatively to other terms involving
gluons. The asymmetry is small but clearly visible at RHIC (the errors
are calculated with $L = 0.8 fb^{-1}$).
On the other hand, a new interaction between quarks could be at the
origin of some deviations from the expected $A_{LL}^{PV} $, provided that
it presents a particular chiral structure.
\vspace{0.5cm}

\no \underline{\it Quark compositeness} \cite{ptjmvct} :
\vspace{0.4cm}

In the framework of compositeness,
an effective handed interaction due to
a contact term between quarks can be present \cite{ELP83} :
\begin{equation}
{\cal L}_{qq} = \epsilon \, {\pi\over {2 \Lambda_{qq}^2}}
\, \bar \Psi \gamma_\mu (1 - \eta \gamma_5) \Psi . \bar \Psi
\gamma^\mu (1 - \eta \gamma_5) \Psi
\end{equation}
\noindent
where $\Psi$ is a quark doublet, $\Lambda$ is the compositeness scale,
$\epsilon$ and $\eta$ taking the values $\pm 1$.

Last year, the CDF collaboration has reported an excess of events
in jet production at high $E_T$ and invariant mass $M_{jj}$
\cite{CDFe}. This can be interpreted as a manifestation of quark
compositeness for a scale $\Lambda \sim 1.6\, TeV$. The
situation is controversial, since D0 doesn't observe such an excess. 
Then we will take this value as the present limit.

\begin{figure}[h]
\vspace{-0.5cm}
    \centerline{\psfig{figure=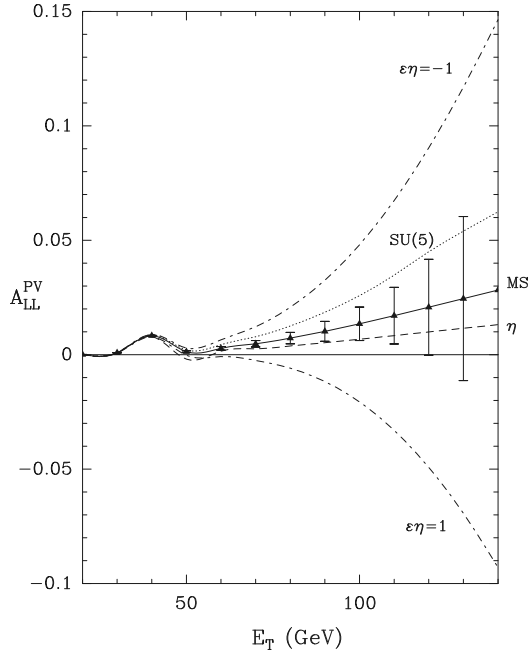,width=8cm}}
\vspace{-1cm}
    \caption[]{Asymmetry $A_{LL}^{PV}$ for inclusive one-jet
production at RHIC. }
    \label{figb2} 
\end{figure}

\no The effects of such contact interactions at RHIC (the scale is 
$\Lambda = 1.6\, TeV$) are represented
by the dot-dashed curves on Fig.2.  The effect
on $A_{LL}^{PV} $ is spectacular. Moreover, if the contact term
is indeed present, this  measurement should allow to
get a unique information on its chiral structure since
the sign of $A_{LL}^{PV} $
is sensitive to the sign  of the product $\epsilon.\eta$
($\epsilon = -1$ means constructive interference, see \cite{ELP83,ptjmvct}).\\
In the absence of deviation from SM expectations,
it will be possible to place some bounds on $\Lambda$. Using a $\chi^2$
analysis, we have obtained the limits given in Table 2. For comparison,
we have also given the bounds achievable from the measurement of
the unpolarized one-jet cross section at the upgrade TEVATRON 
($\sqrt{s} = 2\, TeV$) and at LHC ($\sqrt{s} = 14\, TeV$).

\vspace{0.5cm}

\begin{tabular}{|c|cc|ccc|cc|}
\hline
 & RHIC & & &TEVATRON  & &LHC & \\
$L$ ($fb^{-1}$) & 0.8 & 3.2 & 1. & 10. & 100. & 10. & 100. \\
\hline
$\epsilon = -1$ & 3.30 & 4.40 & 3.20 & 3.70 & 4.10 & 25.5 & 33.0 \\
\hline
$\epsilon = +1$ & 3.25 & 4.35 & 2.90 & 3.35 & 3.75 & 16.0 & 19.5 \\
\hline
\end{tabular} 
\begin{center}
Table 2: Limits on $\Lambda$ in TeV at 95\% CL.
\end{center}

\no It appears that the luminosity is a key parameter for the polarized
analysis on \ALLPV (see the rise of the RHIC bounds with $L$). Finaly,
in spite of its four time lower energy, the RHIC could compete with the
TEVATRON for the observation of compositeness thanks to beam polarization.
\vspace{0.5cm}

\no \underline{\it Leptophobic $Z'$} \cite{ptjmvz} :
\vspace{0.4cm}

 Concerning new neutral gauge bosons ($Z'$), we can show that
there is an equivalence (to a good approximation) between the
exchange of a $Z'$ and a contact term \cite{pcjmv,CTe}. Then,
\ALLPV in the $pp$ mode is also sensitive to the presence of
a new neutral current. We know that the best constraints on
$Z'$ bosons (will) come from the analysis of the Drell-Yan
process. But, a particular $Z'$ could exist, called
"leptophobic", which have zero or very small couplings to
leptons, implying that the traditional $Z'$ searches at future
collider through the leptonic decay channel will be hopeless.
Such leptophobic $Z'$ arise naturaly in some models
derived from string theories \cite{FL}. For illustration we
use the "flipped SU(5)" model(s)
from \cite{LNsu5} and the "$\eta$-kinetic" model from
\cite{BKMR}. A general and interesting property of these
models is that there is parity violation in the up quark sector.
Then, if a leptophobic $Z'$ is present at a relatively low mass,
some effects should be visible on $A_{LL}^{PV}$. We have represented
on Fig.2
the effects of the "flipped SU(5)" model with $M_{Z'}=300\, GeV$
and $\kappa =g_{Z'}/g_Z=1$,
and the "$\eta$-kinetic" model with $M_{Z'}=150\, GeV$ and $\kappa=1$.
Using a $\chi^2$ analysis on \ALLPV with $L=800\, pb^{-1}$, we obtain
the bounds : $M_{Z'_{SU(5)}} > \kappa . 350\, GeV$ and 
$M_{Z'_{{\eta}-k}} > \kappa . 170\, GeV$.
\vspace{0.5cm}

\no \underline{\it $W'$ search} \cite{ptjmvw} :
\vspace{0.4cm}

Finaly, it appears that \ALLPV is also sensitive to the presence of
a new charged current, essentially in the context of polarized
$pn$ collisions. For a $W'_R$ originating from Left-Right Models
\cite{Moh} the present limits depend on several parameters, and, in general,
they are stringent \cite{LS}. But, there is a particular window,
for the right-handed quark mixing matrix $V^R \sim $ {\bf 1} and
for a heavy Dirac right-handed neutrino, where the limits are very
weak \cite{LS}. In this case, \ALLPV measurements at RHIC could
be really interesting, see \cite{ptjmvw} for a detailled analysis.


\end{document}